Silicon-Driven Facet Regulation Enables Tunable Micro-Diamond Architectures in Liquid Ga–In

*Zhi Jiang,[A,B] Xueying Zhang,[A,B] António José Silva Fernandes,[C] Marco Peres,[D,E,F] Gil Gonçalves [A,B] ∗*

A. B. Zhi Jiang, Xueying Zhang, Gil Gonçalves

Centre for Mechanical Technology and Automation (TEMA), University of Aveiro, Aveiro 3810-193, Portugal

Intelligent Systems Associate Laboratory (LASI), Guimarães 4800-058, Portugal

E-mail: (ggconcalve@ua.pt)

C. António José Silva Fernandes

i3N and Physics Department, University of Aveiro, 3810-193 Aveiro, Portugal

D.E.F. Marco Peres

Instituto de Engenharia de Sistemas e Computadores - Microssistemas e Nanotecnologia, Rua Alves Redol 9, 1000-029 Lisboa, Portugal

IPFN, Instituto Superior Técnico, University of Lisbon, Av. Rovisco Pais 1, 1049 001 Lisbon, Portugal

DECN, Instituto Superior Técnico, University of Lisbon, Estrada Nacional 10 (km 139.7), 2695 066 Bobadela, Portugal

Funding: The Horizon Europe (HORIZON), HORIZON-MSCA-2023-PF-01, Grant 101149632–GDSNL; Project UID 00481 Centre for Mechanical Technology and Automation (TEMA); Project CarboNCT, 2022.03596.PTDC (DOI: 10.54499/2022.03596.PTDC)

(We report an ambient-pressure liquid-metal-assisted CVD strategy that enables shape-programmable growth of micro-scale diamond by coupling liquid-metl Ga–In with ferrocene ($Fe(C_5H_5)_2$) as an carbon precursor, nanodiamond seeds, and nanosilicon. Building on liquid-metal diamond synthesis, this approach pushes liquid-metal growth toward a low temperature (900 °C, 1 atm) while enabling single-crystal diamonds to be scaled from ~10 μm to several tens of micrometers with well-developed faceting. Ferrocene decomposition supplies a sustained interfacial carbon flux that is captured and redistributed by the Ga–In melt toward seed-rich liquid–solid interfaces. Defect-rich nanodiamond provides the crystallographic template required for reliable $sp^3$ nucleation despite the intrinsically low carbon solubility of Ga–In. Nanosilicon plays a distinct, complementary role by tuning interfacial kinetics and facet competition, enabling deliberate control of crystal habit: cubic (~10 μm), truncated-tetrahedral, and fully faceted octahedral diamonds are reproducibly obtained by adjusting the nanosilicon:nanodiamond ratio, with octahedral crystals reaching ~50 μm. Importantly, crystal size is further scaled by regulating hydrogen flow: lowering the $H_2$ rate increases net carbon retention at the liquid-metal interface, raises effective supersaturation, and accelerates diamond deposition. Together, habit control (via nanosilicon: nanodiamond) and size scaling (via $H_2$ flow) establish a practical route silicon-driven facet regulation and size under ambient pressure, offering a pathway to tunable micro-sized single-crystal diamonds under mild conditions.)

# 1. Introduction

Over the past few decades, diamond synthesis has progressed from geologically inspired high-pressure techniques to a central topic in materials science [1, 2] and quantum technologies. [3, 4] Traditional high-pressure high-temperature (HPHT) methods, which were developed to reproduce deep-Earth thermodynamic conditions, [5-7] have been refined for improved impurity control, [6] crystallographic quality, [8] and vacancy-center engineering in bulk single crystals. [9, 10] In parallel, microwave plasma chemical vapor deposition (MPCVD) has enabled low-pressure diamond growth with precise isotopic control, [11] tunable doping, [12] and substrate versatility, [13] accelerating the development of diamond films and devices for optoelectronics and quantum information science. [14, 15]

An alternative approach has recently emerged through liquid-metal (LM) mediated diamond synthesis, [16, 17] which aims to induce diamond crystallization under near-ambient pressure and moderate temperatures [18, 19] via mechanisms reminiscent of solution-phase epitaxy. [20, 21] Ruoff and co-workers demonstrated that liquid Ga can decompose carbon precursors to yield 5-10 μm micro-diamonds (MDDs) at ~1025 °C, [19] establishing that low-melting-point metals such as Ga and In can dissolve carbon and facilitate its reprecipitation as $sp^3$ domains. This study introduces a conceptual shift toward pressure-free, plasma-free diamond formation. However, these early low-melting-temperature LM studies were largely limited to a few micrometers' particles, [16, 22] providing little insight into crystallographic orientation, seed-mediated growth, and scalability. Moreover, the mechanistic pathway from carbon accumulation in the melt to the emergence of stable diamond nuclei remains insufficiently resolved, particularly when solid carbon precursors rather than gaseous hydrocarbons are employed. The fundamental sequence governing the $sp^2$-to-$sp^3$ conversion in liquid metals is therefore still poorly understood.

Here, we show that micro-scale single-crystal diamond domains, ranging from 10 to 50 μm, can be synthesized in LM at 900 °C and 1 atm using a solid-state precursor. This approach integrates three major components: ferrocene (FC) acts as both carbon source and Fe-catalytic element, nanodiamond (NDD) seeds providing lattice-matched nucleation sites, and nanosilicon (NSi) serving as an interfacial promoter that modulates surface energies and stabilizes specific facets during growth. During FC decomposition, the released Fe substantially enhances the otherwise limited carbon solubility of Ga–In, enabling the melt to accommodate a higher carbon flux, thereby supporting the formation of micron-scale diamonds. Under these combined influences, porous carbon networks act as a transient $sp^2$- rich scaffold that through hydrogen etching, Fe-mediated reconstruction, and NDD-seeded stabilization—collapse into localized sp³ nuclei, [23] marking the earliest identifiable stage of diamond formation in the LM CVD. [24]

This study explores the seed-assisted diamond formation within a Ga–In liquid-metal environment and deep investigate how the melt conditions facilitate nucleation and facet development under mild conditions. The Ga–In system provides high wettability and moderate carbon diffusivity, producing a confined but chemically active interface, where local carbon supersaturation supports sp³ hybridization. This methodology offers a practical and energy-efficient pathway for ambient-pressure diamond crystallization and enables controlled access to multiple micro-scale diamond morphologies.

## 2. Result and discussion

Different growth conditions were systematically examined, and the average diamond size (ADS) was used to categorize the resulting crystals into three main characteristic morphologies. After refinement of the dataset, three reproducible classes, cubic, truncated tetrahedral, and octahedral, were identified under ambient-pressure (1 atm) LM-CVD, each corresponding to a distinct combination of NSi, NDD, and FC.

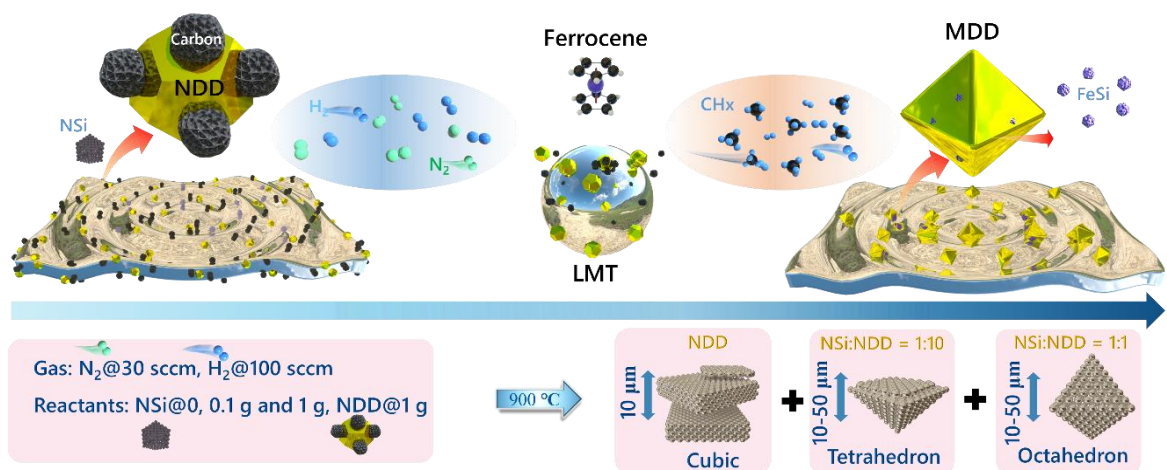

**Figure 1.** Schematic of the LM-assisted CVD pathway for MDD synthesis at 1 atm. Step 1: FC, NDD, and NSi are lightly ground and dispersed in liquid Ga–In. Step 2: At 900 °C under $H_2/N_2$, FC decomposes to $CH_x$ species, supplying carbon to dissolve and accumulate at the LM interface. Step 3: Carbon reconstructs on NDD/NSi-modified interfaces to nucleate and grow diamond; tuning NSi:NDD yields cubic (FC+NDD), truncated-tetrahedral (1:10), and well-faceted octahedral (1:1) MDDs.

Based on systematic experimental observations, we identified three mechanistic pathways that define the LM-assisted CVD growth of micro-sized diamonds (MDDs), as shown in **Figure 1**.

In this system, FC acts as both a solid carbon source and a precursor of Fe catalysts, whereas NDD and NSi function as crystalline seeds and interfacial promoters, respectively.

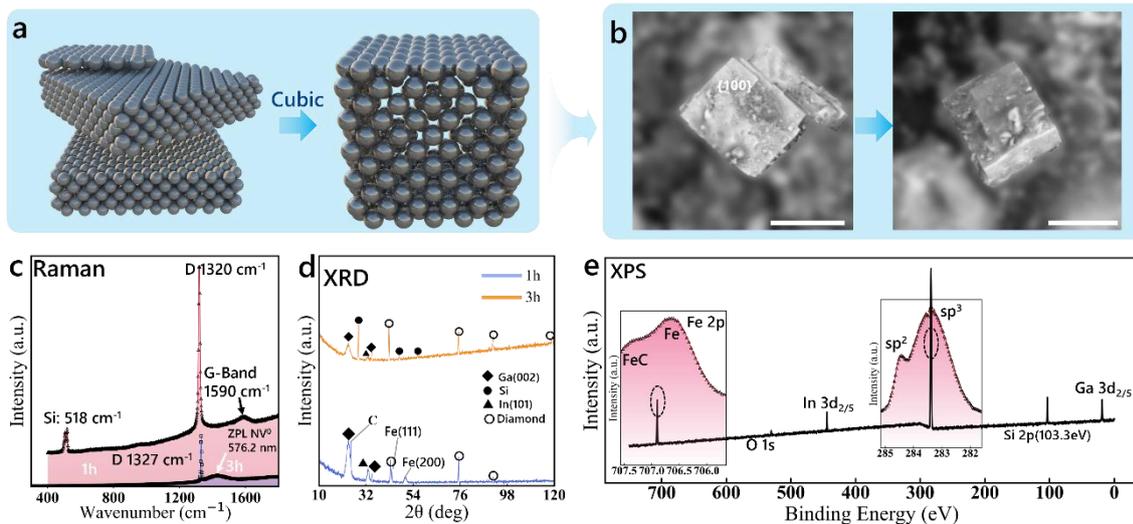

**Figure 2.** After 3 h, Cubic MDDs forms in only NDD in the LM, (a) schematic representation of the process of growth, (b) multilayer deposition of diamond and create 3D structures, (c) Raman evolves from an sp²-rich G-Band (1590 cm⁻¹, 1 h), and NV0 (1430 cm⁻¹, 3 h) to a sharp diamond peak (~1327 cm⁻¹, 3 h). (d) XRD shows the disappearance of Fe–C reflections at 43.9° and 50.3° (2θ). (e) XPS confirms the presence of dominant sp³ carbon (~283 eV) with Fe 2p components at 707.4 eV and 706.6 eV, the strong Ga-In (Ga 3d$_{2/5}$, and In 3d$_{2/5}$) signal indicates that the LM on the surface diamond was not completely removed. Scale bar: 10 μm.

## 2.1. Cubic Morphology

The cubic diamonds produced in the LM system using NDD + FC under N$_2$:H$_2$ = 30 sccm:100 sccm ratio. These ADS-10 μm crystals consist of well-defined square subunits stacked along the (100) direction (**Figure 2a**), which is consistent with the layer-by-layer assembly observed in seeded CVD.

The cubic diamond structure arises from the cooperative action of the Fe species released during FC decomposition and defect-rich NDD (**Figure 2b**). In the LM-assisted CVD environment, the FC provides both the carbon supply and the catalytic Fe needed to activate sp²-rich fragments. Carbon dissolved in the Ga–In melt migrates to the droplet surface, sp²-dominated intermediate formed when carbon first accumulates on the LM surface.[25] As temperature and hydrogen exposure continue, these networks gradually break down: hydrogen removes unstable sp² domains, while the locally increasing carbon concentration—enhanced by Fe-assisted dissolution and redistribution within the LM-assisted carbon supersaturation at the liquid–solid interface. Once the interfacial carbon concentration becomes sufficiently high, sp³ nuclei begin to form within or along the porous framework.

Instead of forming directly, cubic crystals emerge through a recognizable sequence. The earliest deposits consist of a porous carbon network (**Figure S1**), which undergoes hydrogen-assisted thinning and Fe-driven reconstruction. These porous frameworks gradually break down into small faceted blocks, which then reorganize into ~10 μm units composed of stacked (100) layers. Continued carbon addition and facet stabilization allow these cubic subunits to merge into larger microcrystalline diamonds. [26]

The Raman spectra capture the full transformation pathway: at 1 h, the surface is dominated by an sp²-rich carbon film with a pronounced G-Band (~1590 cm$^{-1}$) in **Figure 2c**. Under 532 nm excitation, the NV⁰ ZPL at ~575.9 nm appears at an apparent Raman shift of ~1430 cm$^{-1}$. After 3 h, the stable charge state is determined by the local Fermi level, NV⁻ is favored in electron-rich (n-type) environments, whereas NV⁰ appears when the interface is electron-depleted or surface-oxidized. For instance, when a single carboxyl group replaces one C–OH bond on MDD surface by acid treatment, and deep acceptor states was introduced into the bandgap above the empty NV⁻ defect, [27] resulting in a partial NV⁻-to-NV⁰ conversion. XRD measurements reveal that the Fe–C related reflections at 43.9° and 50.3° (2θ) in **Figure 2d**, visible at early growth, [28] disappear by 3 h, indicating that the Fe-rich intermediate layer becomes buried or consumed as the diamond lattice develops. Complementary XPS results confirm a dominant sp³ C 1s component at ~283.0 eV, along with Fe 2p signals at 707.4 eV (FeC) and 706.6 eV (Fe) in **Figure 2e**. The detected Si 2p (103.3 eV) originates from the underlying Si/SiO$_2$ substrate, consistent with the absence of NSi in the cubic-growth condition. Together, the Raman, XRD, and XPS analyses verify that the cubic particles are fully crystallized sp³ diamonds, with Fe acting only as a transient catalyst during early-stage carbon activation.

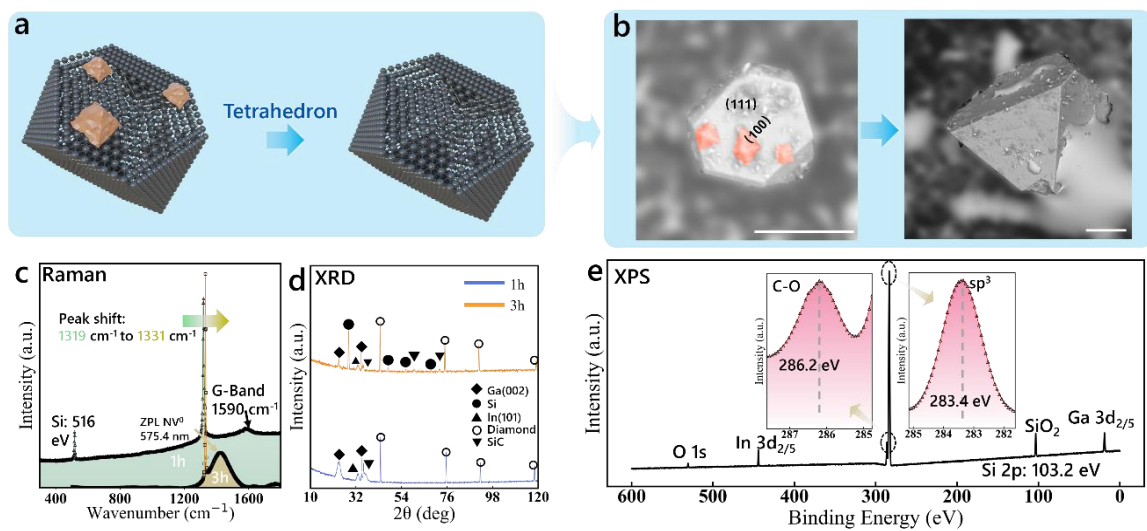

**Figure 3.** After 3 h, truncated tetrahedral MDDs forms in NSi:NDD=1:10, (a) schematic representation of facet truncation anisotropic growth. (b) Small pyramidal protrusions appear on the larger facets, (c) Raman shifts from 1319 to 1331 cm$^{-1}$ and NV$^0$ (1421 cm$^{-1}$, 3 h), indicating stress release during crystallization. (d) Early SiC (35.4°) and Si(111) (28.3°) peaks fade at later stages. (e) XPS shows C–O (286.2 eV) and SiO$_2$ (103.2 eV), and residual LM Ga (3d2/5) and In (3d2/5) are still on the MDD surface.

## 2.2. Truncated tetrahedron Morphology

**Figure 3a** shows the spectroscopic and structural characteristics of the truncated tetrahedral diamonds grown under the NSi:NDD = 1:10 condition. This morphology develops primarily under Si-assisted growth, where NSi enhances carbon incorporation and promotes rapid facet evolution (**Figure 3b**). At a weight ratio of NSi:NDD = 1:10, truncated tetrahedra become the dominant product, although a few octahedral grains may still appear under these conditions. The enhanced interfacial reactivity introduced by the NSi increases carbon mobility within the LM, enabling faster propagation along selected crystallographic directions [29] and producing crystals up to ~50 μm in size.

As the Si concentration increases toward which critical regime, the liquid–solid interface becomes chemically heterogeneous. [19] Transient Si–C and Si–O surface species alter the effective surface energies of different planes, producing asymmetric attachment kinetics between the (111) and (100) facets. [30] The formation of this structure when local supersaturation or impurity-driven diffusion enhances the growth rate of vicinal facets relative to the parent plane. [31] In some grains, small pyramidal protrusions appear on the larger facets, suggesting that the truncated tetrahedron is not solely a product of surface erosion but may emerge from continued anisotropic growth once facet stability becomes imbalanced. [32] Taken together, these observations indicate that the truncated tetrahedral MDD represents an intermediate, dynamically evolving morphology arising from the competition between Si-driven enhancement, directional carbon attachment, and mild LM-mediated reconstruction. Under 532 nm excitation, the NV⁰ ZPL at ~575.4 nm appears at an apparent Raman shift of ~1421 cm⁻¹, Raman spectroscopy captures the progressive crystallization process: the diamond band shifts from 1319 cm⁻¹ at 1 h in **Figure 3c**, indicative of strained or partially ordered sp³ nuclei to 1331 cm⁻¹ at 3 h, [33] where the line shape becomes sharper. This evolution matches the well-known trajectory from amorphous carbon through nanocrystalline intermediates to fully developed diamonds, driven by stress relaxation and reduced phonon confinement during lattice ordering. [34]

XRD analysis shows that a weak SiC reflection at 35.4° (2θ) appears during early deposition but disappears in **Figure 3d** with continued growth, [35] suggesting that Si-containing species act transiently at the interface rather than forming stable carbide phases. A small Si (111) peak at 28.3° (2θ) persists, in agreement with the Si–Si feature (~99.3 eV) observed by XPS in **Figure 3e**. These signatures indicate that NSi acts as a promoter, modifying the interfacial chemistry, but not being incorporated into the diamond lattice. The XPS C 1s spectrum shows a minor C–O component (~286.3 eV), [36] likely generated during the mixed-acid cleaning process. Collectively, the Raman, XRD, and XPS data confirm that the truncated tetrahedral crystals consist of well-crystallized sp³ diamond whose morphological evolution reflects Si-mediated anisotropic growth.

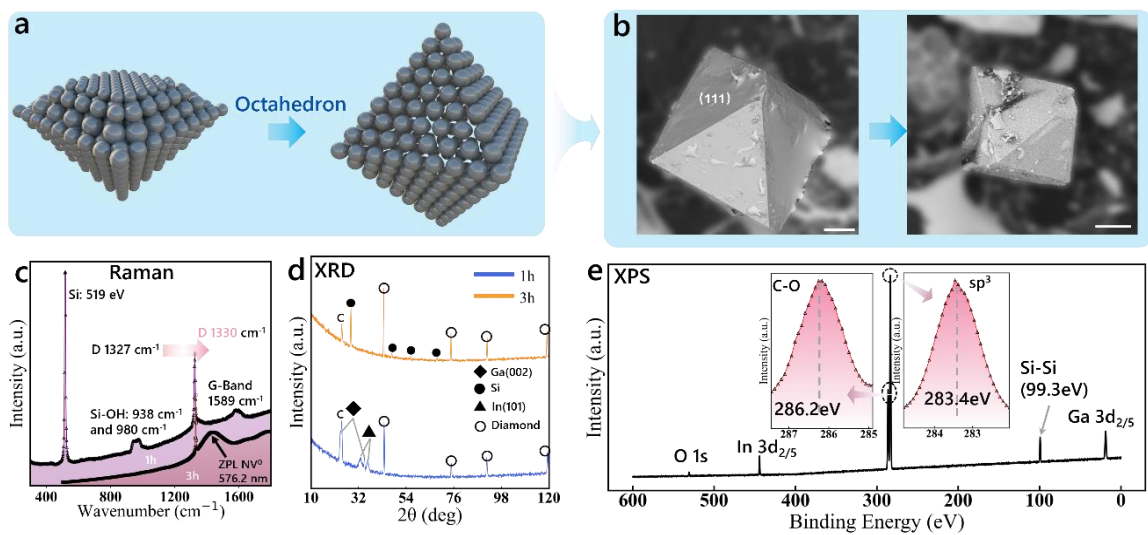

**Figure 4.** After 3 h, Octahedral MDDs forms in NSi:NDD=1:1, (a) schematic representation of (111) isotropic growth, (b) Si enrichment promotes balanced (111) growth, and occasional pyramidal intermediates reflect incomplete facet closure, (c) Raman exhibits diamond peaks at 1327–1330 cm$^{-1}$ with weak Si–OH features (938–980 cm$^{-1}$) and NV$^0$ (1435 cm$^{-1}$, 3 h). (d) XRD reveals diamond reflections with a minor Si(111) contribution. (e) XPS displays C–O signatures like those in truncated tetrahedral MDDs, C–O component (~286.2 eV) formed after acid treatment.

## 2.3. Octahedron Morphology

When the NSi:NDD ratio is increased up to 1:1, the majority of MDDs evolve into well-defined octahedral crystals (**Figure 4a**), representing the most thermodynamically favored morphology under these growth conditions. The higher abundance of Si markedly enhances the interfacial carbon mobility and homogenizes the surface energy within the LM medium, enabling more uniform growth across the (111) facets. [37] This stabilization of (111)-terminated surfaces promotes the formation of complete octahedra, in sharp contrast to the truncated tetrahedral morphology that dominates at a lower NSi:NDD ratio (1:10). As Si-containing species become progressively immobilized at the interface, the system approaches a compositional steady state, allowing the facets to evolve more smoothly and symmetrically. This interplay results in the fully faceted sp³ octahedral diamonds observed in **Figure 4b**. This behavior aligns with prior studies showing that Si additives facilitate the transition from kinetically biased to thermodynamically dominated diamond growth by enhancing interfacial diffusion and promoting facet equilibration.

The NV⁰ ZPL at ~576.2 nm appears at an apparent Raman shift of ~1435 cm$^{-1}$ (@ 532 nm). Raman spectra display two additional, weak features near 938 cm$^{-1}$ and 980 cm$^{-1}$ in **Figure 4c**, attributed to Si–OH vibrations that likely originate from partial oxidation of Si-rich interfacial residues followed by acid cleaning. The main diamond band shifts from 1327 cm$^{-1}$ (1 h) to 1330 cm$^{-1}$ (3 h), following the same crystallization trend observed in **Figure 3c**, but with narrower line widths, consistent with more complete ordering of the sp³ framework and reduced internal strain in the fully faceted octahedra. This relaxation of lattice distortion, together with improved crystallinity, manifests as a stronger and sharper diamond peak in both Raman and XRD, especially for truncated tetrahedron and Octahedron MDDs. Correspondingly, the diamond reflections exhibit increased intensity and reduced FWHM, consistent with grain coarsening, diminished strain, and the transition from early-stage porous frameworks to fully faceted sp³ microcrystals. [38] XRD patterns show a faint Si (111) reflection, indicating that NSi remains at the interface between the LM and MDD in **Figure 4d**. This is consistent with the XPS results, where the C 1s peak is dominated by sp³ carbon (~283 eV) in **Figure 4e**, accompanied by a small C–O component similar to that in the truncated tetrahedral sample. Importantly, Fe–C related reflections observed at earlier stages vanish entirely, while the characteristic diamond peaks at 2θ ≈ 43°, 75°, 91°, and 119° become well-defined in, reflecting high crystallinity.

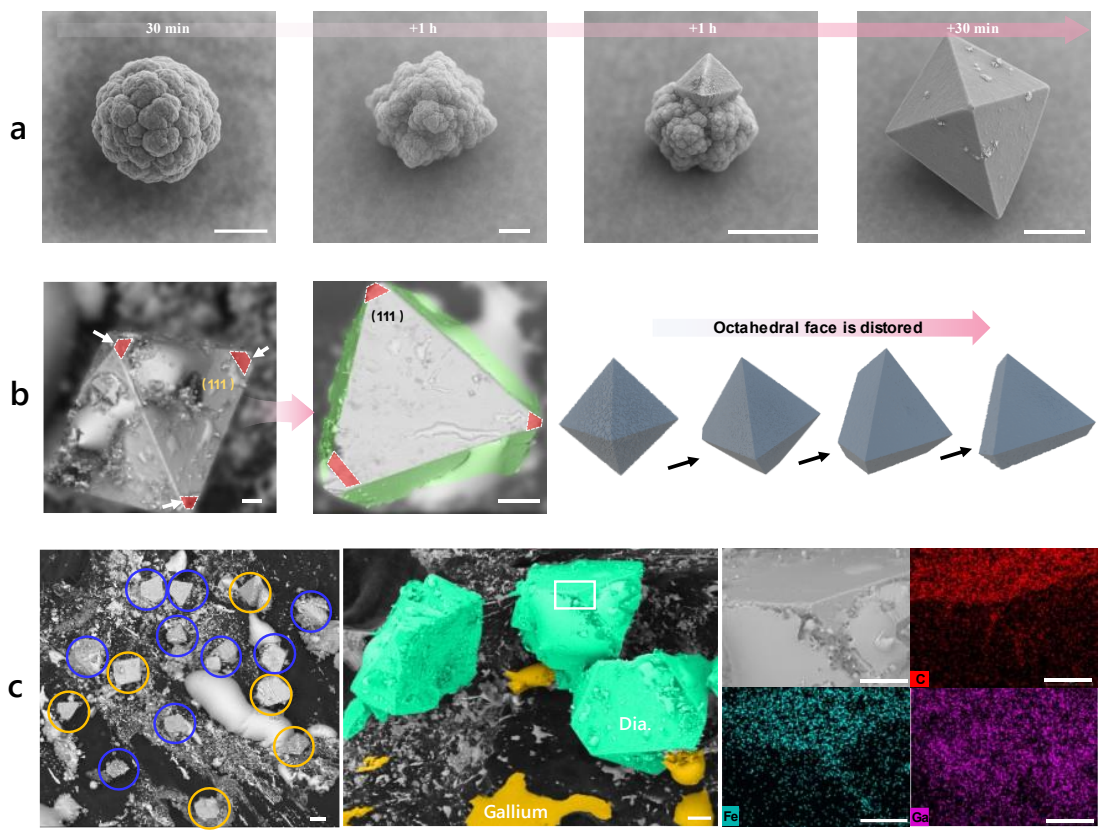

**Figure 5.** Effect of NSi content on diamond morphology in the LM Ga–In–FC system. (a) Si-content modulation using NSi:NDD = 1:10 (fully embedded in Ga–In before CVD). (b) Low Si (NSi:NDD = 1:10) yields truncated-tetrahedral diamonds due to unbalanced growth between (111) and (100) facets. (c) Higher Si (NSi:NDD = 1:1) promotes more complete (111) faceting; truncated-tetrahedral and octahedral crystals coexist, indicating a shift toward more balanced growth. Scale bar: 10 μm.

## 2.4. Effect of NSi:NDD Ratio on Facet Kinetics and Diamond Habit Evolution

We discussed the formation pathways of the three characteristic diamond habits (cubic, truncated-tetrahedral, and octahedral) and verified the sp³ diamond phase by Raman, XRD, and XPS. To place these observations on a consistent basis, it is necessary to address a common, practical issue in liquid-metal growth: after cooling, Ga–In can remain as residual films or patches on diamond surfaces, partially obscuring the true faceting and introducing additional signals in EDS/XPS. Such interfacial residues can complicate direct comparisons across different NSi:NDD conditions and growth times. Therefore, we next perform a comparative analysis using standardized post-treatment and control observations to clarify how surface metal retention evolves, how it affects apparent morphology and surface chemistry, and how reliably the intrinsic diamond facets can be evaluated after cleaning.

As shown in **Figure 5a**, the SEM image presents diamond grains after post-treatment in sulfuric acid and nitric acid followed by deionized-water rinsing. Ga-In adheres strongly to diamond surfaces after LM growth due to hydrophobic trapping that form upon exposure to trace oxygen during cooling. Similar residual Ga layers have been reported in LM diamond and graphene growth, where high wettability and interfacial oxides impede spontaneous detachment. To remove LM Ga-In, the crystals were treated in 50 °C $HNO_3:H_2SO_4$ (1:1), a standard oxidizing mixture that converts Ga and Ga–oxide layers into soluble $Ga^{3+}$ species. Subsequent DI-water ultrasonication removes remaining fragments, yielding clean faceted diamond surfaces. [19, 39] In contrast, the earlier samples (≤ 2.5 h growth) were not subjected to this cleaning step, as they consisted of porous carbonaceous diamond precursors with inherently hydrophobic surfaces, resulting in only minor LM retention. The effect of NSi concentration on diamond morphology was investigated using two precursor systems: (i) one containing NSi: NDD = 1:10, and a (ii) comparative sample with NSi: NDD = 1:1. In both cases, excess LM ensured uniform coverage, and during high-temperature CVD, the solid particles migrated to the Ga–In droplet surfaces, where crystallization and nucleation produced MDDs [40] following the pathway shown in **Figure 1**. EDS analysis (**Figure S2**) confirms that only trace Fe and Si remain on the diamond surfaces, while carbon dominates the bulk, forming well-faceted octahedral crystals. A minor oxygen signal likely originates from thin Fe/Si-oxide interfacial layers formed by residual air in the CVD chamber that facilitates Ga-In wetting or from carboxyl (–COOH) groups generated during acid cleaning. These results indicate that acid treatment efficiently eliminated metallic residues, allowing an accurate assessment of the intrinsic morphology and purity of the octahedral diamonds.

In **Figure 5b** (NSi:NDD = 1:10), the limited Si supply places the system in a kinetics-dominated growth regime. [19] At this low concentration, Si at the LM interface is insufficient to homogenize surface energies or balance the relative growth rates of the (111) and (100) facets. As a result, the early octahedral embryos develop facet-dependent propagation, where discrepancies in the advancement of (111) and (100) planes cause gradual distortion of the evolving crystal. This imbalance drives selective truncation of vertex regions, yielding the characteristic truncated-tetrahedral geometry composed of two large triangular faces, three smaller triangular facets, and three quadrilateral planes. Transient species such as Si–C and Si–O formed during the process should also affect the growth kinetics of MDDs, contributing to asymmetric growth. [41] [30]

In **Figure 5c** (NSi:NDD = 1:1), the diamond grains collected from the LM surface exhibit both truncated-tetrahedral (yellow) and octahedral (blue) morphologies in roughly comparable numbers. The higher Si content enriches the LM interface, improving carbon mobility and interfacial wetting while partially equalizing the growth rates of the (111) and (100) facets. This enhanced diffusion–wetting environment stabilizes the (111) surfaces and increases the likelihood of forming complete octahedra, although truncated tetrahedra remain prevalent owing to ongoing facet competition during growth. SEM and EDS (**Figure S3**) analyses confirm that some MDDs retain thin, patch-like Ga-In residues; given the hydrophobic nature of diamond, such metallic films adhere primarily through Si- and Fe-oxide interfacial layers formed during deposition and cooling, compared without acid-treated MDD (**Figure S4**). The coexistence of both morphologies at NSi:NDD = 1:1 underscores the delicate balance between Si-assisted surface homogenization and the intrinsic facet kinetics of diamond growth.

Time-resolved SEM/EDS reveals a clear interfacial evolution during LM-CVD. After 0.5 h, a porous, carbon-rich (sp²-dominated) network forms on Ga–In droplets and traps small LM residues, serving as a transient scaffold prior to sp³ nucleation (**Figure S5a**). By 1.5 h, carbon enrichment increases (**Figure S6**) while Ga/Fe/Si signals decline, and diamond nuclei emerge with rough, incompletely stabilized facets (**Figure S5b**). At 2.5 h, the porous framework reorganizes into layered diamond domains with only trace residual metal (**Figure S5c**), indicating progressive lattice ordering and carbon overwhelmingly dominates the composition (**Figure S7a-S7c**). With continued growth to 3 h (**Figure S8**), these domains develop into well-faceted microdiamonds, where minor surface Ga/Fe/Si can locally reconstruct edges into multi-faceted polyhedral forms.

Collectively, the comparative experiments in **Figure S9** clarify three practical factors that govern LM-CVD microdiamond formation and help rationalize the growth trends discussed above. First, reducing the H$_2$ flow (from N$_2$:H$_2$ = 30:100 to 30:30 sccm) increases net carbon retention at the Ga–In interface, leading to a higher nucleation density (**Figure S10**) and enabling growth of well-faceted crystals (**Figure S9a**) from the ~10 μm scale up to >50 μm (**Figure S11**). Second, carbon accumulation and graphite boat on the liquid metal is not sufficient for diamond formation (**Figure S9b**): in the absence of both NDD and NSi, the products remain sp²-rich (EDS and Raman characterization of graphene/graphite-like, as shown in **Figure S12**) despite the presence of the graphite boat, indicating that the observed microdiamonds arise from the seeded, promoter-assisted pathway rather than from boat-derived carbon alone. Third, changing the substrate (**Figure S9c**) and precursor configuration confirms the robustness of the interfacial mechanism while highlighting how Si distribution influences kinetics: diamonds reproducibly form on both the graphite-boat configuration and on a Si wafer with NDD, and the time-resolved evolution from porous carbon scaffolds to faceted microcrystals supports the view that NDD establishes reliable sp³ nucleation, whereas Si primarily modulates interfacial conditions and facet competition during subsequent growth.

## 3. Conclusion

Micro-scale faceted diamonds (cubic, truncated-tetrahedral, and octahedral) were synthesized at 1 atm in a Ga–In liquid-metal-assisted CVD system at a reduced growth temperature of 900 °C by introducing FC as the solid carbon precursor together with NDD seeds and NSi promoters. The results clarify that morphology is governed primarily by the NSi:NDD ratio: NDD-only conditions favor predominantly ~10 µm cubic crystals dominated by (100) growth, NSi:NDD = 1:10 drives kinetically unbalanced facet competition and yields truncated-tetrahedral habits, and increasing to NSi:NDD = 1:1 enhances interfacial regulation and (111) stabilization, producing a much higher fraction of well-faceted octahedra (often coexisting with truncated forms) and enabling sizes to extend from ~10 µm to several tens of micrometers. In addition to habit control, hydrogen flow provides an independent handle on size scaling by modulating net carbon retention at the liquid–solid interface, which supports the formation of crystals exceeding ~50 µm under lower $H_2$ flow. Control experiments further indicate that carbon accumulation from the graphite boat alone does not lead to sp³ diamond formation under otherwise identical conditions, whereas diamond growth appears only when FC and the NDD/NSi additives are present, supporting FC as the effective carbon source in this system. Finally, substrate comparisons (graphite boat versus Si wafer) show that the physical state and distribution of the liquid metal and seeds on different substrates alter droplet formation and local interfacial environments, which in turn influences nucleation density, growth evolution, and the resulting diamond habit.

## 4. Experimental Section
### 4.1. Methods

Diamond growth experiments were performed in a horizontal ceramic tube CVD system (Termolab, TH123) operated under ambient pressure (1 atm). Solid ferrocene (Fe(C$_5$H$_5$)$_2$, 98%, Sigma-Aldrich) was used as the carbon source and placed in a graphite boat at the center of the furnace. The growth zone, located at the center of the furnace, contained the LM Ga-In eutectic alloy (95:5 weight ratio, 99.99%), into which either NDD or NSi seeds were directly added. All seed materials were used as received, without further purification or surface treatment. The NDD seeds (210 nm, >99% purity) and NSi (50 nm, Alfa Aesar) were uniformly dispersed within the LM Ga-In prior to heating.

A premixed N$_2$/H$_2$ gas flow (30 sccm: 100 sccm) was introduced through mass flow controllers (Termolab, TH123) and continuously maintained throughout the experiment, as shown in **Figure 6**. After initial purging (15 min), the furnace temperature was ramped to 900 °C at 10 °C/min and held isothermally for 3 h under constant gas flow. During this period, FC (boiling point - 249 °C) was transported via the carrier gas into the high-temperature growth zone, where carbon species were absorbed into the LM and underwent recrystallization.

Control experiments were performed using NDD seeds and/or NSi while maintaining all other parameters identical. Upon completion, the furnace was allowed to cool naturally to room temperature under a flowing nitrogen atmosphere. The solidified products were collected from the alloy surface and subjected to structural, morphological, and compositional analyses.

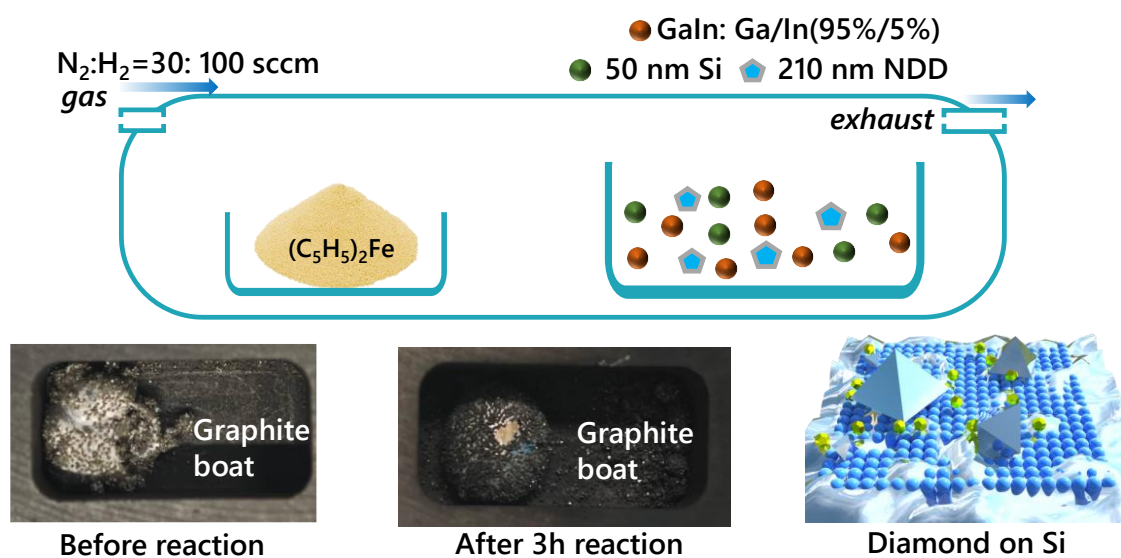

**Figure 6.** Schematic of the diamond growth setup: the FC on the left decomposes to supply carbon species, whereas the NSi and NDD seeds are dispersed in the LM Ga-In. MDD grows on Si wafer. Optical images of the graphite boat before and after the CVD process.

## 4.2. Characterization

Raman spectroscopy and SEM were employed to characterize the bonding structure and surface morphology of the synthesized micro-scale diamond crystals. After growth, the solidified products were retrieved from the Ga-In melted and rinsed sequentially with ethanol and dilute hydrochloric acid (10%) to remove most of the residual surface-bound metals. The samples were then dried at 60 °C before analysis.

Raman spectra were acquired using a Renishaw in Via confocal Raman microscope equipped with a 532 nm excitation laser, under ambient conditions. Spectra were collected in the range of 500–2000 cm$^{-1}$ with a spectral resolution of ~10 cm$^{-1}$. A prominent first-order diamond peak at ~1332 cm$^{-1}$ was consistently observed after the CVD process, confirming the presence of sp$^3$-hybridized carbon characteristic of a crystalline diamond. The sharpness and intensity of the 1332 cm$^{-1}$ band indicate the good crystallinity of the diamond phase. In some spectra, weak and broad bands centered around ~1440 cm$^{-1}$ (D-band) were also detected, suggesting minor amounts of NV centers, possibly originating from NDD. No significant Raman signals from the gallium or tin compounds were observed in the tested range.

SEM imaging was performed using a Zeiss Gemini SEM 500 field-emission scanning electron microscope operated at 15 kV. Secondary electron images revealed faceted diamond crystallites with well-defined morphologies, typically ranging from 10 to 100 μm in size. The crystals displayed smooth type faces and sharp edges, indicative of epitaxial growth. Occasionally, metallic remnants were observed adhering to the crystal surfaces, especially in samples with incomplete post-growth acid treatment. EDS was also used to map the residual Ga, confirming its peripheral presence and absence from diamond core regions.

The XPS spectra of the as-prepared MDDs were obtained using a Kratos AXISULTRA DLD scanning electron microscope. Absorption was measured by using a Lambda 1050 UV/vis/NIR spectrophotometer.

These analyses confirmed the successful formation of micron-sized diamonds with characteristic sp$^3$ carbon bonding and provided direct evidence that seed-mediated diamond crystallization can occur under ambient pressure in an LM environment.

**Acknowledgements**

The authors thank António Fernandes for their kind help with the measurements. The authors acknowledge the financial support from The Horizon Europe (HORIZON), HORIZON-MSCA-2023-PF-01, Grant 101149632–GDSNL, project UID 00481 Centre for Mechanical

Technology and Automation (TEMA) and project CarboNCT, 2022.03596.PTDC (DOI: 10.54499/2022.03596.PTDC) are acknowledged. ChatGPT was used solely to assist with English-language editing and improving clarity; all scientific content, data interpretation, and conclusions were verified by the authors.

**Conflicts of Interest**

The authors declare no conflict of interest.

**Data Availability Statement**

Supplementary information: The online version contains supplementary material available at Correspondence should be addressed to Gil Goncalves

**References**

1. Ekimov, E. A.; Lyapin, S. G.; Boldyrev, K. N.; Kondrin, M. V.; Khmelnitskiy, R.; Gavva, V. A.; Kotereva, T. V.; Popova, M. N., *JETP Letters* **2015,** *102* (11), 701-706.
2. Yan, B.; Jia, X.; Sun, S.; Zhou, Z.; Fang, C.; Chen, N.; Li, Y.; Li, Y.; Ma, H.-a., *International Journal of Refractory Metals and Hard Materials* **2015,** *48*, 56-60.
3. Aghaeimeibodi, S.; Vuckovic, J. In *Quantum Photonics with SnV Centers in Diamond*, 2021 IEEE International Electron Devices Meeting (IEDM), San Francisco, CA, USA, 2021-12-16; IEEE: San Francisco, CA, USA, **2021**; pp 14.6.1-14.6.4.
4. Debroux, R.; Michaels, C. P.; Purser, C. M.; Wan, N.; Trusheim, M. E.; Arjona Martínez, J.; Parker, R. A.; Stramma, A. M.; Chen, K. C.; de Santis, L.; Alexeev, E. M.; Ferrari, A. C.; Englund, D.; Gangloff, D. A.; Atatüre, M., *Physical Review X* **2021,** *11* (4), 041041.
5. Fattaruso, L., *Physics Today* **2024,** *77* (7), 14-16.
6. Dossa, S. S.; Ponomarev, I.; Feigelson, B. N.; Hainke, M.; Kranert, C.; Friedrich, J.; Derby, J. J., *Journal of Crystal Growth* **2023,** *609*, 127150.
7. Plakhotnik, T.; Duka, T.; Davydov, V. A.; Agafonov, V. N., *Diamond and Related Materials* **2023,** *139*, 110363.
8. Sonin, V.; Tomilenko, A.; Zhimulev, E.; Bul'bak, T.; Chepurov, A.; Babich, Y.; Logvinova, A.; Timina, T. y.; Chepurov, A., *Scientific Reports* **2022,** *12* (1), 1246.
9. Razgulov, A. A.; Lyapin, S. G.; Novikov, A. P.; Ekimov, E. A., *Diamond and Related Materials* **2021,** *116*, 108379.
10. Zhang, X.; Liu, K.-Y.; Li, F.; Liu, X.; Duan, S.; Wang, J.-N.; Liu, G.-Q.; Pan, X.-Y.; Chen, X.; Zhang, P.; Ma, Y.; Chen, C., *Advanced Functional Materials* **2023,** *33* (52), 2309586.

11. Kawano, M.; Hirama, K.; Taniyasu, Y.; Kumakura, K., *Journal of Applied Physics* **2024,** *135* (7), 075302.
12. Yusuke Tominaga, Y. M. H., Naoaki Kubota, Naoya Ishida, Susumu Sato, Takeshi Kondo, Makoto Yuasa, Hiroshi Uetsuka, Chiaki Terashima, *Diamond & Related Materials* **2023,** *140* (110543).
13. Deng, Z.; Liu, C.; Jin, Y.; Pu, J.; Wang, B.; Chen, J., *The Analyst* **2019,** *144* (15), 4569-4574.
14. Santos, N. E.; Mendes, J. C.; Braga, S. S., *Molecules (Basel, Switzerland)* **2023,** *28* (4).
15. K. C, A.; Siddique, A.; Anderson, J.; Saha, R.; Gautam, C.; Ayala, A.; Engdahl, C.; Holtz, M. W.; Piner, E. L., *SN Applied Sciences* **2022,** *4* (8), 226.
16. Wang, M.; Lin, Y., *Nanoscale* **2024,** *16* (14), 6915-6933.
17. Zhao, Z.; Soni, S.; Lee, T.; Nijhuis, C. A.; Xiang, D., *Advanced Materials* **2023,** *35* (1), 2203391.
18. Wu, X.; Fang, H.; Ma, X.; Yan, S., *Advanced Optical Materials* **2023,** *11* (22), 2301180.
19. Yan Gong, D. L., Myeonggi Choe, Yongchul Kim, Babu Ram, Mohammad Zafari, Won Kyung Seong, Pavel Bakharev, Meihui Wang, In Kee Park, Seulyi Lee, Tae Joo Shin, Zonghoon Lee, Geunsik Lee, Rodney S. Ruoff, *Nature* **2024,** *629*, 348–354.
20. Guo, S.; Ji, Y.; Liao, G.; Wang, J.; Shen, Z.-H.; Qi, X.; Liebscher, C.; Cheng, N.; Ren, L.; Ge, B., *Journal of the American Chemical Society* **2024,** *146* (29), 19800-19808.
21. Wu, K.; Dou, Z.; Deng, S.; Wu, D.; Zhang, B.; Yang, H.; Li, R.; Lei, C.; Zhang, Y.; Fu, Q.; Yu, G., *Nature Nanotechnology* **2024**.
22. Li, L.; Zhang, Q.; Geng, D.; Meng, H.; Hu, W., *Chemical Society Reviews* **2024,** *53* (13), 7158–7201.
23. Prabowo, J.; Lai, L.; Chivers, B.; Burke, D.; Dinh, A. H.; Ye, L.; Wang, Y.; Wang, Y.; Wei, L.; Chen, Y., *Carbon* **2024,** *216*, 118507.
24. Silva, T. A.; Zanin, H.; May, P. W.; Corat, E. J.; Fatibello-Filho, O., *ACS Applied Materials & Interfaces* **2014,** *6* (23), 21086-21092.
25. Sedelnikova, O. V.; Gorodetskiy, D. V.; Lavrov, A. N.; Grebenkina, M. A.; Fedorenko, A. D.; Bulusheva, L. G.; Okotrub, A. V., *Synthetic Metals* **2024,** *307*, 117675.
26. Yang, G.; Sun, P.; Zhu, T.; Wang, Y.; Li, S.; Liu, C.; Yang, G.; Yang, K.; Yang, X.; Lian, W.; Peng, Z.; Lu, Y.; Liu, H.; Jiang, N., *Diamond and Related Materials* **2024,** *141*, 110600.
27. Kaviani, M.; Deák, P.; Aradi, B.; Frauenheim, T.; Chou, J.-P.; Gali, A., *Nano Letters* **2014,** *14* (8), 4772-4777.
28. Xiao-Hong, F.; Bin, X.; Zhen, N.; Tong-Guang, Z.; Bin, T., *Chinese Physics Letters* **2012,** *29* (4), 048102.
29. Khokhryakov, A. F.; Palyanov, Y. N.; Borzdov, Y. M.; Kozhukhov, A. S.; Sheglov, D. V., *Diamond and Related Materials* **2018,** *87*, 27-34.
30. Gogotsi, Y.; Welz, S.; Ersoy, D. A.; McNallan, M. J., *Nature* **2001,** *411* (6835), 283-287.
31. Wang, Y.; Song, C.; Han, S.; Wang, Z.; Zhang, X.; Hu, X.; Ge, L.; Xu, M.; Peng, Y.; Wang, X.; Hu, X.; Xu, X., *Vacuum* **2025,** *240*, 114554.
32. Zhang, T.; Sun, F.; Wang, Y.; Li, Y.; Wang, J.; Wang, Z.; Li, K. H.; Zhu, Y.; Wang, Q.; Shao, L.; Wong, N.; Lei, D.; Lin, Y.; Chu, Z., *ACS Nano* **2024,** *18* (52), 35405-35417.
33. Ruiz-Valdez, C. F.; Chernov, V.; Meléndrez, R.; Álvarez-García, S.; Santacruz-Gómez, K.; Berman-Mendoza, D.; Barboza-Flores, M., *physica status solidi (a)* **2018,** *215* (22), 1800267.

34. Stuart, S. A.; Prawer, S.; Weiser, P. S., *Diamond and Related Materials* **1993,** *2* (5), 753-757.
35. Arnault, J. C.; Saada, S.; Delclos, S.; Intiso, L.; Tranchant, N.; Polini, R.; Bergonzo, P., *Diamond and Related Materials* **2007,** *16* (4), 690-694.
36. Wang, X.; Ruslinda, A. R.; Ishiyama, Y.; Ishii, Y.; Kawarada, H., *Diamond and Related Materials* **2011,** *20* (10), 1319-1324.
37. Xie, Z. Q.; Bai, J.; Zhou, Y. S.; Gao, Y.; Park, J.; Guillemet, T.; Jiang, L.; Zeng, X. C.; Lu, Y. F., *Scientific Reports* **2014,** *4* (1), 4581.
38. NAGASHIMA, K.; NARA, M.; MATSUDA, J.-i., *Meteoritics & Planetary Science* **2012,** *47* (11), 1728-1737.
39. Dickey, M. D., *ACS applied materials & interfaces* **2014,** *6* (21), 18369-18379.
40. Luo, D., *ECS Meeting Abstracts* **2025,** *MA2025-01* (13), 1064.
41. Pantea, C.; Voronin, G. A.; Waldek Zerda, T.; Zhang, J.; Wang, L.; Wang, Y.; Uchida, T.; Zhao, Y., *Diamond and Related Materials* **2005,** *14* (10), 1611-1615.

**Supporting Information**

Supporting Information is available from the Chemrxiv or from the author.

**Ambient-pressure Ga–In liquid-metal CVD converts ferrocene-derived carbon into faceted microdiamonds at 900 °C. Nanodiamond seeds define nucleation, nanosilicon tunes facet kinetics to switch between cubic, truncated-tetrahedral, and octahedral habits, and reduced H₂ flow increases carbon retention to grow crystals beyond 50 μm. Control runs show graphite alone yields only sp² carbon, and substrates influence morphology of micro-scale single crystal diamond.**

**TOC**

**Ambient-pressure Ga–In liquid-metal CVD converts ferrocene-derived carbon into faceted microdiamonds**

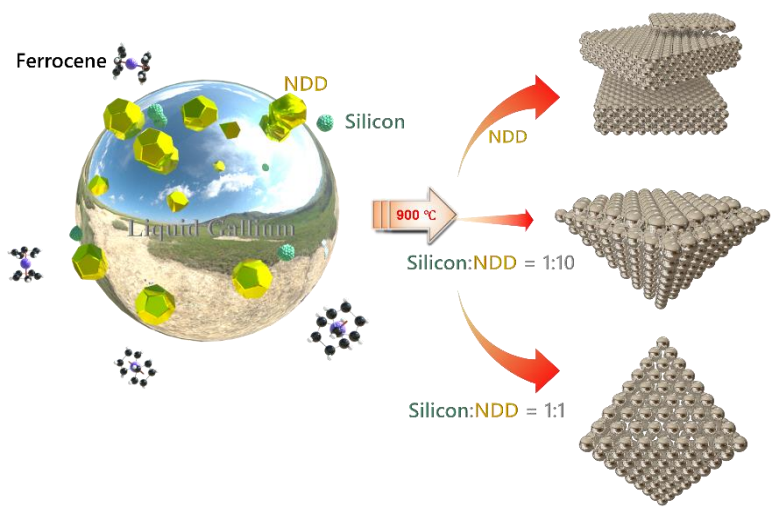